\documentclass[12pt,preprint]{aastex}

\shorttitle{Hypercompact H {\small II} Region G28.20$-$0.04N}
\shortauthors{Sewi{\l}o et al.}

\begin{document}

\title{Internal Dynamics of the Hypercompact H {\small II} Region G28.20$-$0.04N}

\author{M. Sewi{\l}o\altaffilmark{1}, E. Churchwell\altaffilmark{2}, S. Kurtz\altaffilmark{3}, 
W. M. Goss\altaffilmark{4}, and  P. Hofner\altaffilmark{4,5}} 
\altaffiltext{1}{Space Telescope Science Institute, 3700 San Martin Drive, Baltimore, MD 21218, mmsewilo@stsci.edu}
\altaffiltext{2}{University of Wisconsin - Madison, Department of Astronomy, 475 N. Charter St., 
Madison, WI 53706, churchwell@astro.wisc.edu}
\altaffiltext{3}{Centro de Radioastronom\'\i a y Astrof\'\i sica, Universidad Nacional Aut\'onoma
de M\'exico, Apdo. Postal 3-72, 58089, Morelia, Mich. M\'exico, s.kurtz@astrosmo.unam.mx}
\altaffiltext{4}{National Radio Astronomy Observatory, P.O. Box 0, Socorro, NM 87801, mgoss@aoc.nrao.edu}
\altaffiltext{5}{New Mexico Tech, Physics Department, 801 Leroy Place, Socorro, NM 87801, phofner@nrao.edu}

\begin{abstract}

High resolution (0$\rlap.{''}$15) Very Large Array observations of 7 mm continuum and H53$\alpha$ line emission toward the hypercompact H {\small II} region G28.20-0.04N reveal the presence of large-scale ordered motions. We find a velocity gradient of 10$^{3}$ km s$^{-1}$ pc$^{-1}$ along the minor axis of the continuum source. Lower resolution (1$\rlap.{''}$0--2$\rlap.{''}$3) radio recombination line observations indicate a systematic increase of line width from H30$\alpha$ to H92$\alpha$. Under the assumption that the H30$\alpha$ line does not suffer significant pressure broadening, we have deconvolved the contributions of statistical broadening (thermal, turbulent, and pressure) from large-scale motions.  The pressure broadening of the H53$\alpha$, H76$\alpha$, and H92$\alpha$ lines implies an electron density of 6.9$\times$10$^{6}$ cm$^{-3}$, 8.5$\times$10$^{5}$ cm$^{-3}$,  and 2.8$\times$10$^{5}$ cm$^{-3}$, respectively.

\end{abstract}


\section{Introduction}

Hypercompact (HC) H {\small II} regions have about one tenth the size ($<$0.05
pc) and about a hundred times the density (n$_{e}>$10$^{5}$
cm$^{-3}$) of ultracompact (UC) H {\small II} regions.  It has been postulated that HC H IIs represent the transition from hot molecular cores to UC H II regions \citep{ku00}; that is, the transition from rapid accretion onto a massive central protostar
to the termination of accretion and the emergence of a detectable radio H II region. If HC H {\small II}s occupy
this evolutionary niche, they should show evidence of ordered large-scale motions such as expansion, infall, bipolar
outflows, rotation and shocks due to accretion and outflows. Also, owing to their high electron densities, recombination lines from the H$^{+}$ gas should suffer substantial pressure broadening in high quantum state radio recombination lines (RRLs).

It is not clear if HC H {\small II} regions are simply smaller and
denser versions of UC H {\small II} regions or if they are 
intrinsically different.  Their relative
sizes, however, suggest that stellar clusters may be responsible
for UC H {\small II} regions, while HC H {\small II} regions may be 
produced by a single, or perhaps a binary, OB protostar.
Some, but not all, HC H {\small II}s have unusually broad radio recombination
lines (BRRLs) with full widths at half maximum intensity (FWHM) $>$40
km s$^{-1}$.  The typical FWHM of more evolved (UC and compact) H {\small II}
regions is $\sim$25-30 km s$^{-1}$ in all RRLs from principal quantum
states n$<$100. 
The fact that not all HC H {\small II}s have BRRLs may imply that BRRLs are only
present for a limited period during the earliest HC H {\small II} evolutionary
phase when ordered large-scale motions are maximum.  

\citet[and in prep.]{s04} observed the H92$\alpha$,
H76$\alpha$, and H53$\alpha$ lines toward the HC H {\small II} region
G28.20-0.04N (hereafter G28.20N) with resolutions of $2\rlap.{''}3$,
$1\rlap.{''}5$, and $1\rlap.{''}6$, respectively, all of which are
larger than the angular size of the HC H {\small II} region.  They found FWHMs
of 74 km s$^{-1}$, 58 km s$^{-1}$, and 40 km s$^{-1}$, respectively.
The rapid change in line width with transition strongly suggests
that pressure broadening is a dominant broadening mechanism in
G28.20N.

We have chosen G28.20N for a high resolution H53$\alpha$ study to
determine the origin of the BRRLs in this source because: it is a
bona fide HC H {\small II} region; it has already been observed in several RRLs
at lower angular resolution; several masers have been detected toward
G28.20N implying extreme youth (H$_{2}$O - \citealt{h98}; OH 6035
MHz - \citealt{cv95}; OH 1667 MHz - \citealt{a00}; and,
CH$_{3}$OH - \citealt{m91}). The H53$\alpha$ line was chosen because it
should be mostly free from pressure broadening, the nebula should be optically thin 
at this frequency, and the Very Large Array can achieve an angular resolution one tenth that of previous 
observations at 43~GHz. 

\section{Observations}

G28.20N was observed with the VLA\footnote{The Very Large Array (VLA) is operated by the National Radio Astronomy Observatory (NRAO), a facility of the National Science Foundation operated under cooperative agreement by Associated Universities, Inc.} at 7 mm (continuum and H53$\alpha$ line) with an angular resolution of
$0\rlap.{''}15$. The observations were made on 2004 January 19 with the VLA in its B-configuration.
At the H53$\alpha$ rest frequency of 42.95197 GHz, this provided an angular
resolution of  0$\rlap.{''}$17 $\times$ 0$\rlap.{''}$13 at a position angle of $-4\rlap.^{\circ}$8.
A 25~MHz bandwidth was employed, with 31 spectral line channels of 781~kHz each.  This provided
a velocity resolution of 5.5~km~s$^{-1}$, which was later Hanning smoothed to 11~km~s$^{-1}$.
 Integration time on-source was about 1.4 hours. 
3C 286 (1.47 Jy), 1851$+$005 (0.68$\pm$0.01 Jy), and 3C 345 (4.81$\pm$0.18 Jy) were used as flux density, phase and bandpass calibrators, respectively. The data reduction, including self-calibration, followed the VLA high-frequency guidelines for calibration and imaging, using the NRAO package AIPS. The continuum image is shown in Figure 1.
Continuum-subtracted line images were used to generate H53$\alpha$ maps. 

\section{Results}

\subsection{Morphology of the Source}

At our high angular resolution, G28.20N appears as a shell-like
structure, extended in the SE-NW direction (Figure 1).  There are two
peaks, separated by $0\rlap.''36$ (2050 AU or 0.01 pc at a distance of
5.7 kpc; \citealt{f03}). The brightness distributions along one-dimensional cuts through
the source show a well-defined shell with low intensity at the center.
The center of the shell-like nebula is at RA = 18$^{\rm h}$42$^{\rm m}$58$\rlap.^{\rm s}$11 and Dec = $-$04$^{\circ}$13$'$57$\rlap.''$4 (J2000). The source diameter at 10\% of the peak (the 12$\sigma$ contour in Figure 1) is $\theta_{s}=0\rlap.''9$ (5100 AU;
where $\theta_{s}=(\theta_{RA} \cdot \theta_{Dec})^{1/2}$). The
integrated flux density is 645$\pm$65 mJy, where the 10\%
uncertainty reflects VLA calibration uncertainty in the 7~mm band. The
synthesized beam brightness temperature using the peak brightness of 64 mJy
beam$^{-1}$ (located at the W side of the shell) is 1920 K, implying a
peak optical depth of 0.3 based on T$_{e}$ = 7000 K, and a peak emission
measure (EM) of 1.6$\times$10$^{9}$ pc cm$^{-6}$.  This EM implies an rms
electron density of $n_e$ = 6.2$\times$10$^5$~cm$^{-3}$.
The synthesized beam brightness of the central cavity, 11 mJy~beam$^{-1}$,
corresponds to 343 K, implying an optical depth of 0.05, an EM of 
2.5$\times$10$^{8}$ pc cm$^{-6}$, and an rms $n_{e}$ of 2.5$\times$10$^{5}$~cm$^{-3}$.

\subsection{Velocity Structure}

We examined the velocity structure of G28.20N by making contour plots
of the H53$\alpha$ emission, using
moment maps, position-velocity diagrams, and by Gaussian fitting of
the line profiles at multiple positions. All these methods reveal a
velocity gradient of 10$^{3}$ km s$^{-1}$ pc$^{-1}$, with velocities
increasing from the NE to SW.

In Figure 2 we show least-squares Gaussian fits to the H53$\alpha$
line integrated over different areas of the source, each approximately one synthesized beam in size.
The positions are indicated in the continuum image of Figure 2. The
boxes are located along the major and minor axes of the shell, with
two boxes (B and D) sampling the regions of maximum emission. Box C
covers the low intensity central region, while boxes F and I trace an
extension in the SE--NW direction. We also analyzed the
line emission integrated over the entire source (Box J). The
 parameters of the Gaussian fits are given in Table 1; $V_{LSR}$ is the velocity relative to the local standard of rest, FWHM$_{deconv}$ is the line FWHM corrected for instrumental broadening, $S_{L}$ is the flux density in excess of the continuum at line center, and $S_{C}$ is the continuum flux density. In some directions (e.g. E and H) 
the line profiles are asymmetric with an additional velocity component apparent. All the asymmetries 
occur in the velocity range of $\sim$110-130 km s$^{-1}$. Although weak, these features 
are real; we also detect a 130 km s$^{-1}$ component in the H76$\alpha$ line. 

The sequence of five H53$\alpha$ line profiles shown in Figure 2
indicates a systematic shift in the central line velocity from the NE to
SW. The dotted vertical line indicates the 88.2 km~s$^{-1}$ systemic
velocity of the nebula obtained from H53$\alpha$ data taken with 10
times lower resolution than shown in boxes A-E ({\it upper right};
HPBW $\sim$ 1.$''$6)\footnote{Our 7 mm flux density and H53$\alpha$ line width (both B- and D-array) disagree with the results of \citet{k08} for their 7 mm/H53$\alpha$ data (S$_{\rm int}$=1.1 Jy and FWHM = 33.4$\pm$1.5 km s$^{-1}$). To investigate this discrepancy, we re-reduced our lower resolution 7 mm data and the archival Keto et al. data (AK607) using consistent procedures for both data sets.  We confirmed our results and found that the Keto et al. data are consistent with our own. For the AK607 data we find an integrated 7 mm continuum flux density of 790$\pm$35~mJy and a FWHM for the H53$\alpha$ line of $38.7\pm0.9$~km~s$^{-1}$.}. Thus, the centroid velocity of $\sim$78 km
s$^{-1}$ in Box A (NE side of G28.20N) is {\it blueshifted} and the
centroid velocity of $\sim$97 km s$^{-1}$ in Box E (SW side of
G28.20N) is {\it redshifted} with respect to the systemic nebular
velocity. The H53$\alpha$ lines suggest rotational motion with the
axis of rotation in the SE--NW direction. There is no velocity
pattern along the elongated SE--NW axis (i.e. no indication of an outflow along this axis). We suggest that the excess line broadening along this axis may be due to an outflow in the plane of the sky, expanding perpendicular to its axis of flow with a velocity of $\sim$5 km s$^{-1}$.

The velocity structure observed in G28.20N is consistent with a torus rotating with a 
velocity of 5.0 km~s$^{-1}$ at a radius of 0.005
pc (1030 AU; at the position of the peaks), which corresponds to a
velocity gradient of about 10$^{3}$ km s$^{-1}$ pc$^{-1}$ or an interior
mass of 28 M$_\odot$ for Keplerian motion.

\subsection{Optical Depth Effects}

We compared integrated line profiles for all transitions and found that the H30$\alpha$ line velocity shifts by about 5-14 km~s$^{-1}$ with respect to the H92$\alpha$, H76$\alpha$, and H53$\alpha$ lines.  The H92$\alpha$, H76$\alpha$, and H53$\alpha$ line velocity integrated over the nebula is 82.9$\pm$0.9 km s$^{-1}$, 78.9$\pm$0.8 km s$^{-1}$, and 88.2$\pm$0.4 km s$^{-1}$, respectively. \citet{k08} reports the H30$\alpha$ line velocity of 92.5$\pm$0.2 km s$^{-1}$. A similar trend of increasing RRL velocity with increasing frequency has been reported for other compact H {\small II} regions, e.g., W49A/B \citep{dmg97} and W3(OH) \citep{be83}.  \citet{wm87} and \citet{k95}
proposed that this is an optical depth effect, in which deeper layers with different velocities are probed at higher frequencies. Following their arguments, the optical depth decreases rapidly at high
frequencies because the absorption coefficient is inversely
proportional to the square of the frequency; thus the H92$\alpha$, H76$\alpha$, H53$\alpha$, and H30$\alpha$ 
emission arise from different volumes of gas with H30$\alpha$ emission being more sensitive to the denser regions. 
They also argue that the velocity of the high frequency RRLs provides a good approximation to the average
motion of the H {\small II} region provided that the random motions are
the same for all densities throughout the nebula.  The velocity of this
innermost H {\small II} gas is expected to most closely indicate the
stellar velocity; hence, for G28.20N, the H30$\alpha$ line velocity
should provide the best estimate of the central star's radial
velocity.  The blueshifted velocities of the outer H {\small II}
region, traced by the H92$\alpha$, H76$\alpha$, and H53$\alpha$ lines, imply
expansion, consistent with a strongly over-pressured nebula.

\section{Line Broadening Mechanisms: Pressure Broadening vs. Dynamic Motions}

\citet{k08} detected the H30$\alpha$ line toward G28.20N using the SMA with an 
angular resolution of 1$''$. The width of the H30$\alpha$ line is 20.9$\pm$0.6 km s$^{-1}$.
At the high frequency of the H30$\alpha$ line, line width should be dominated by the thermal, turbulent, 
and large scale motions; pressure broadening is negligible due to the rapid decrease in pressure broadening with frequency. Thermal and turbulent broadening produce convolved Gaussian profiles whose center velocities generally coincide. Large-scale coherent motions can produce
velocity-shifted profiles that may deviate from Gaussian shapes.  The
observed profiles depend on the type and angular extent of the
large-scale velocity components relative to the synthesized beam.  In
particular, as the angular resolution is increased, each resolution
element should approach the combined thermal plus turbulent broadening
of the gas, with its central velocity reflecting the dominant
large-scale motion of the gas in that volume element.  In some cases,
two components of large-scale coherent motions may fall within a
single beam, which may result in a double Gaussian.  An example of
this is an expanding bubble whose front and back faces are included in
all observations or if the red and blue lobes of an outflow are
projected onto the same resolution element.

Molecular line observations of UC H {\small II} regions and massive cloud cores suggest that 
the typical molecular line width where HC H {\small II} regions form is $\sim$5 km s$^{-1}$  
(FWHM; \citealt{c91}, \citealt{p97}). We adopt this value for the non-thermal 
(turbulent) contribution to the line width since the thermal width for molecules is $<$0.5 km s$^{-1}$. 
For 7000 K, the thermal contribution to the line width is 17.9 km s$^{-1}$ (FWHM). Thus, the combined 
contribution (added in quadrature) of thermal and turbulent widths to a RRL profile is 18.6 km s$^{-1}$. 
We can determine the contribution from kinematic motions by subtracting this value in quadrature from 
the observed H30$\alpha$ line width. The residual after accounting for thermal and turbulent motion is 
9.6 km s$^{-1}$. This excess represents the approximate magnitude of the large-scale motions (detected 
in the H53$\alpha$ line) within G28.20N as the data is averaged over the entire source. For non-thermal widths of 0 km s$^{-1}$ (no turbulent motions) and 10 km s$^{-1}$, the contribution from large-scale motions would be 10.8 km s$^{-1}$ and 4.1 km s$^{-1}$, respectively. The value of the non-thermal contribution to the line width is not essential to the following calculations of pressure broadening. 

The line width in km s$^{-1}$ arising from thermal and turbulent motions is constant for all relevent transitions (from H30$\alpha$ to H92$\alpha$). Thus, we can determine the influence of internal kinematics on the H53$\alpha$, H76$\alpha$, and H92$\alpha$ line widths. The subtraction in quadrature of the 9.6 km~s$^{-1}$ kinematic broadening from the observed line widths (39.7$\pm$1.3, 57.6$\pm$2.2, and 74.4$\pm$2.6 km~s$^{-1}$) gives FWHMs of 38.5, 56.8, and 73.8 km s$^{-1}$ for the H53$\alpha$, 
H76$\alpha$, and H92$\alpha$ lines, respectively; these line widths, ``corrected'' for kinematic
broadening, still contain a pressure broadening contribution.

To estimate the pressure broadening of the H53$\alpha$, H76$\alpha$, and H92$\alpha$
lines, we use an approximation for the Voigt profile width \citep[eq. 2.72]{gs03}, 
which relates the frequency widths of the Gaussian, Lorentz and Voigt profiles.  The Gaussian width for low
frequency lines is determined from the width of a high frequency line,
unaffected by pressure broadening. Here we use the 16.167~MHz
H30$\alpha$ line width.  
For the H53$\alpha$ line, the Gaussian width is 2.99~MHz and the observed
line width is 5.69~MHz, implying a pressure broadening width of
4.03 MHz (or 28.1 km s$^{-1}$). 
For the H76$\alpha$ line, the Gaussian width is 1.02~MHz and the observed
line width is 2.82 MHz, implying a pressure broadening width of
2.43~MHz (or 49.5 km s$^{-1}$). 
For the H92$\alpha$ line, the Gaussian width is 0.58~MHz and the observed 
line width is 2.06 MHz, implying a pressure broadening width of 1.89~MHz 
(or 68.1 km s$^{-1}$). 

The pressure broadening width is proportional to n$^{7.4}$ T$_{e}^{-0.1}$ n$_{e}$; thus, the pressure
broadened lines can be used to determine the true space electron density n$_{e}$ averaged over the 
nebula \citep{bs72}. Using T$_{e}$ of 7000 K, we
find  n$_{e}$ of (6.9$\pm$0.4)$\times$10$^{6}$ cm$^{-3}$, (8.5$\pm$0.4)$\times$10$^{5}$ cm$^{-3}$ and (2.8$\pm$0.1)$\times$10$^{5}$ cm$^{-3}$ based on the H53$\alpha$, H76$\alpha$, and H92$\alpha$
lines, respectively. 
The rms electron densities inferred from the 0.7, 2,
and 3.6 cm continuum observations are 
1.7$\times$10$^{5}$ cm$^{-3}$, 
9.3$\times$10$^{4}$ cm$^{-3}$ \citep[and in prep.]{s04},
and 7.6$\times$10$^{4}$ cm$^{-3}$ \citep{s04},
respectively. The higher electron density from pressure broadening
is expected because it represents the actual space density of the gas
that produces the broad lines as opposed to the rms density derived
from the radio continuum. The fact that the average density decreases with higher quantum transitions 
also suggests that each  line in the range 0.7 to 3.6 cm samples different volumes of gas, as also suggested by the change in line velocity with transition.

The rms and true electron densities can be used to calculate the filling
factor $f$ for the nebula.  Equating the Str\"omgren relation for
a uniform density region (in terms of n$_{rms}$) with that of a clumpy 
region (in terms of n$_{true}$) which is formed of N smaller clumps, one 
obtains  $N r^3_{clump}/r^3_{uniform} = n^2_{e,rms}/n^2_{e,true} = f$.
This is the ratio of the volume occupied by the dense gas to
the total volume of the H {\small II} region; thus $f$ is a measure of the degree of
clumpiness of the nebula. Based on the 7 mm continuum and line data, we
calculated the filling factor of 6$\times$10$^{-4}$ for the highest 
density gas we sample in G28.20N.

\section{Summary}

We have resolved the morphology and velocity structure of the HC H
{\small II} region G28.20N. The continuum emission
 has a shell-like structure, with an inner radius of 1100 AU, an outer radius of 2500 AU,
 and an rms density contrast of 2-3 between the central cavity and the torus.

Under the assumption that the H30$\alpha$ line is not significantly
pressure broadened and that turbulence broadening is $\sim$5 km
s$^{-1}$, we have resolved the contributions of thermal, turbulent,
and pressure broadening from large-scale ordered motions. We suggest that the ordered motion 
has two components: a velocity gradient of 10$^{3}$ km s$^{-1}$ pc$^{-1}$ in the NE--SW direction along the minor axis which may indicate rotation of a torus around a 28~M$_\odot$ object; and line
broadening (line splitting at some points) along the major axis which
may indicate lateral expansion of an outflow along the major axis of the nebula. 
In this scenario, the outflow axis would have to be in the plane of the sky
 because line velocities along this axis are centered on the systemic
 velocity of the nebula. We suggest that the excess line broadening along 
 the major axis is due to expansion of the outflow perpendicular to its axis. 

The line width from large scale motions averaged over the whole nebula
is $\sim$10 km s$^{-1}$. We find pressure broadening line widths for
the H53$\alpha$, H76$\alpha$, and H92$\alpha$ lines of $\sim$28, $\sim$50, and $\sim$68 km
s$^{-1}$, respectively. These spatially unresolved, pressure broadened
lines were used to determine electron densities averaged over the  
nebula (not rms densities). The mean n$_{e}$ is 6.9$\times$10$^{6}$ cm$^{-3}$, 8.5$\times$10$^{5}$ cm$^{-3}$,
and 2.8$\times$10$^{5}$ cm$^{-3}$ based on the H53$\alpha$, H76$\alpha$, and
H92$\alpha$ lines, respectively. These densities are 4--40 times higher
than the rms densities derived from continuum data. The fact that the  
average electron densities decrease with higher quantum transitions is 
consistent with the change in opacity with frequency.

This study has for the first time resolved the contributions of
thermal, turbulent, pressure, and ordered large-scale motions of
BRRLs in a hypercompact H {\small II} region. The observations show
that large scale dynamics and high densities both contribute to BRRLs
in G28.20N. This study was possible only because of the high angular
resolution and sensitivity provided by the VLA.

%\begin{acknowledgements} 
\acknowledgements We thank Michael Burton, the referee, for useful suggestions that improved 
our manuscript. M. S. and E. B. C. acknowledge partial support from NSF grant AST-0303689.
S. K. acknowledges partial support from DGAPA, UNAM grant IN106107. 
%\end{acknowledgements}

%\clearpage

\clearpage

%%%%%%%%%%%%%%%%%%%%%%%%%%%%%%%%%%%%%%%%%%%%%%%%%%%%%%%%%%%%%%%%%%%%%%%%%%%%%%%
% TABLE 1 - Gaussian Fits to the RRLs in Boxes
%%%%%%%%%%%%%%%%%%%%%%%%%%%%%%%%%%%%%%%%%%%%%%%%%%%%%%%%%%%%%%%%%%%%%%%%%%%%%%%
\begin{deluxetable}{lrcrr}
\tablecaption{Gaussian Fits to the High Resolution H53$\alpha$ RRLs\tablenotemark{a}}
\tablewidth{0pt}
\tablehead{
\colhead{Box} &
\colhead{V$_{\rm LSR}$} &
\colhead{FWHM$_{\rm deconv}$}&
\colhead{S$_{L}$}&
\colhead{S$_{C}$} \\
\colhead{} &
\colhead{(km~s$^{-1}$)} &
\colhead{(km~s$^{-1}$)}&
\colhead{(mJy)} &
\colhead{(mJy)} }
\startdata
A & 77.6 $\pm$ 1.7 & 34.9 $\pm$ 4.9 & 5.1 $\pm$ 1.0 & 12.1 $\pm$ 0.6  \\
B & 85.8 $\pm$ 0.5 & 33.8 $\pm$ 1.3 & 22.5 $\pm$ 1.1 & 53.4 $\pm$ 0.6  \\
C & 92.1 $\pm$ 0.7 & 35.9 $\pm$ 2.2 &  6.7 $\pm$ 0.6 & 23.2 $\pm$ 0.6  \\
D & 95.8 $\pm$ 0.4 & 31.6 $\pm$ 1.0 & 26.4 $\pm$ 1.1 & 66.5 $\pm$ 0.6 \\
E\tablenotemark{b} & 97.3 $\pm$ 1.0 & 28.9 $\pm$ 3.0 & 7.8 $\pm$ 1.0 & 17.9 $\pm$ 0.6  \\
\nodata  &\raisebox{1.5ex}[0pt]{\Big \{}127.9 $\pm$ 3.0 & 9.4 $\pm$ 5.0 & 1.8 $\pm$ 1.3& \nodata \\
F & 89.9 $\pm$ 1.3 & 29.6 $\pm$ 3.7 & 4.0 $\pm$ 0.7 & 10.4 $\pm$ 0.6  \\
G & 87.7 $\pm$ 0.8 & 36.6 $\pm$ 2.4 & 11.4 $\pm$ 1.0 & 29.5 $\pm$ 0.6  \\
H\tablenotemark{b} & 82.8 $\pm$ 1.7 & 26.5 $\pm$ 3.3 & 10.1 $\pm$ 1.7 & 34.7 $\pm$ 0.6  \\
\nodata  &\raisebox{1.5ex}[0pt]{\Big \{}110.3 $\pm$ 4.0 & 21.1 $\pm$ 7.4 & 3.6 $\pm$ 1.8 & \nodata \\
I & 83.2 $\pm$ 2.1 & 28.1 $\pm$ 5.9 & 3.4 $\pm$ 0.9 & 8.5 $\pm$ 0.6  \\
\cline{1-5}
J & 89.5 $\pm$ 0.6 & 39.8 $\pm$ 1.7 & 238 $\pm$ 14 & 641 $\pm$ 4   
\enddata
\tablenotetext{a}{Spectra were Hanning smoothed to a resolution of 11.0 km s$^{-1}$.
 Uncertainties are the formal 1$\sigma$ (68.3\% confidence level) deviations of the 
 data from the Gaussian fits.}
\tablenotetext{b}{Two Gaussian components were required to adequately fit the observed 
line profile.}
\end{deluxetable}

\clearpage 

%----------------------------------------------------------
% FIGURE 1: G28.20 - continuum image and slices
%----------------------------------------------------------
\begin{figure}
\plotone{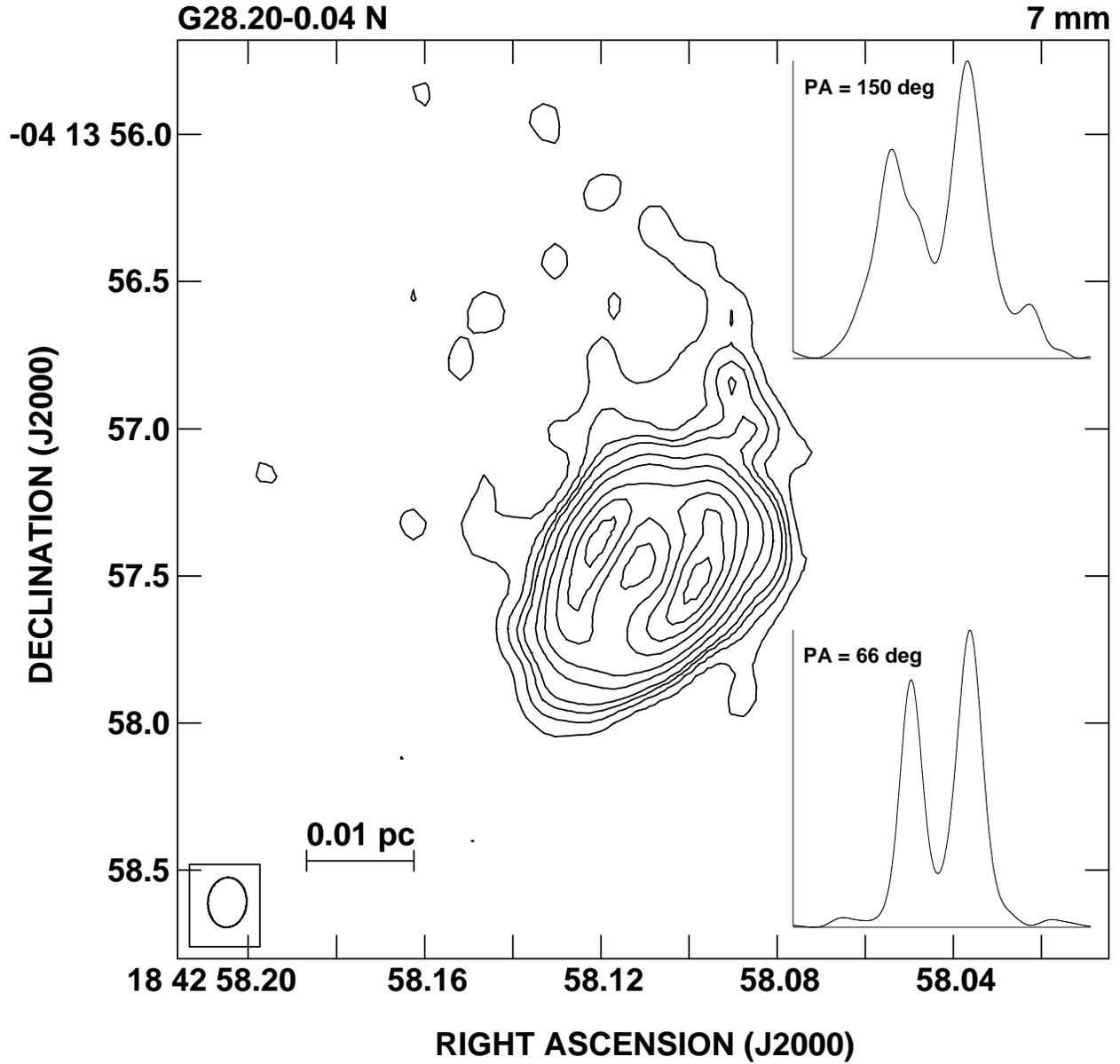}
\caption{The 7 mm image of G28.20$-$0.04 N. 
  The continuum intensity peak is 64.0 mJy~beam$^{-1}$. The
  contours are $-$3, 3, 6, 9, 12, 20, 30, 50, 70, 90, 110 $\times$ 0.53
  mJy~beam$^{-1}$ (the image rms). The synthesized beam of $0\rlap.{''}17$
  $\times 0\rlap.{''}13$ is shown in the lower left corner.
  The insets show intensity cuts through
  G28.20N; position angles indicate rotation East from North.}
\end{figure}

\clearpage

%----------------------------------------------------------
% FIGURE 2: The H53a Line Velocity Gradient
%----------------------------------------------------------
 \begin{figure}
\plotone{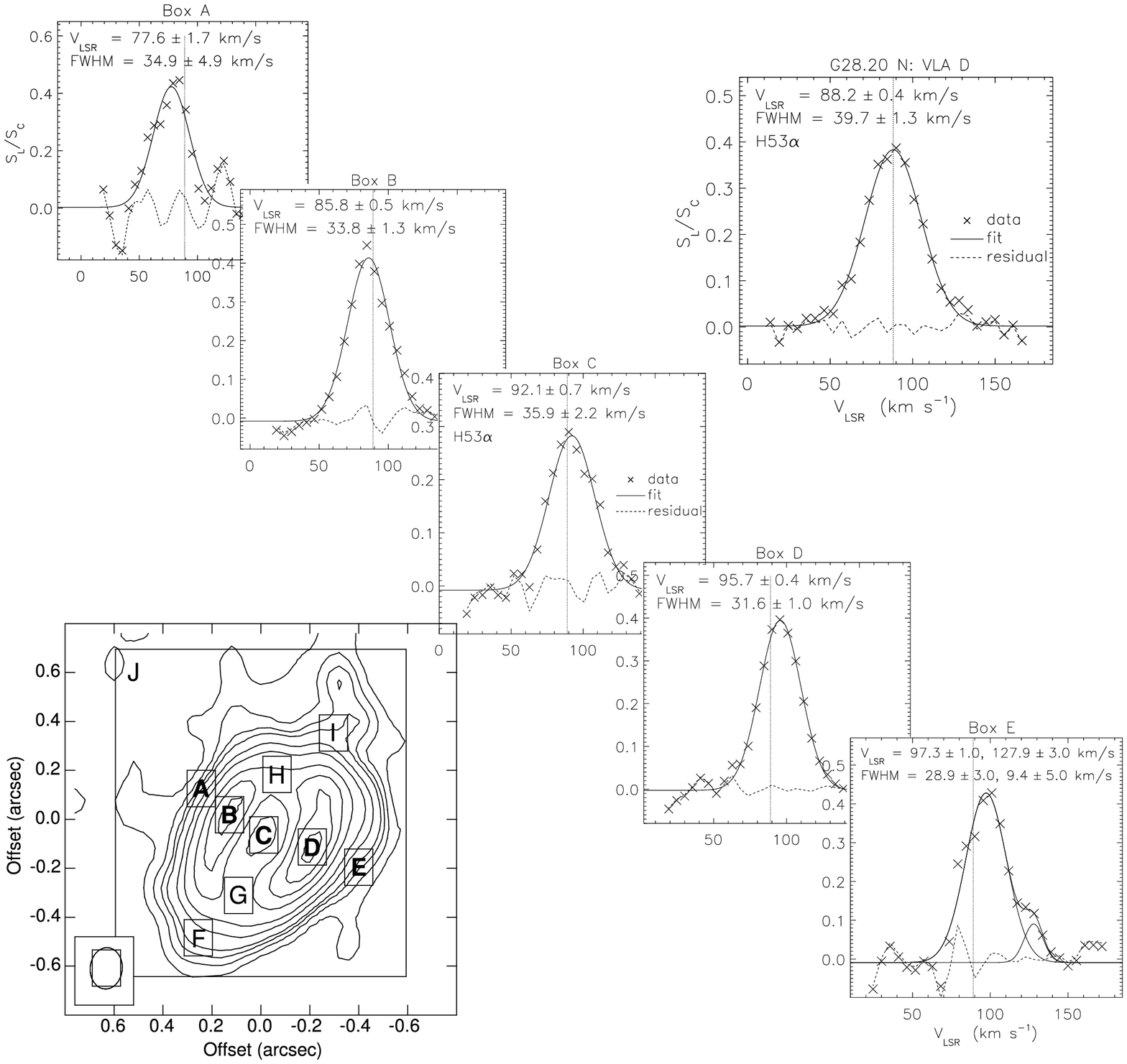}
\caption{A sequence of five H53$\alpha$ line profiles shown in the
  diagonal direction illustrates a shift in the line central
  velocity of $\sim$20 km~s$^{-1}$ along the NE--SW direction. 
  The spectra were Hanning smoothed to a resolution of 11.0 km s$^{-1}$.
  The dotted vertical line indicates the systemic velocity of the
  H53$\alpha$ line profile integrated over the entire nebula from data
  with 10 times lower resolution ({\it upper right}; HPBW $\sim$ $1\rlap.{''}6$, VLA D-Array; S$_{\rm int}$ = 711 mJy). The lines were integrated over rectangular
  areas indicated in the 7 mm continuum image ({\it lower left}). The
  corresponding box letters are given at the top of each panel. In the
  lower left corner of the continuum image, the size of the
  integration box is compared with the synthesized beam of $0\rlap.{''}17
  \times 0\rlap.{''}13$; the two have very nearly the same area.}
\end{figure}

\end{document}